\documentclass[aip,apl,reprint]{revtex4-1}

\usepackage[utf8]{inputenc}
\usepackage{graphicx}
\usepackage{amsmath}
\usepackage{lipsum}
\usepackage{amsfonts}
\usepackage{amssymb}
\usepackage{dcolumn}
\usepackage{bm}
\usepackage{blindtext}
\usepackage{textgreek}
\newcommand{\um}{{\textmu}m }
\newcommand{\uum}{{\textmu}m}

\begin{document}

	\title{Imaging of acoustic pressure modes in opto-mechano-fluidic resonators with a single particle probe}
	
	\author{Jeewon Suh, Kewen Han, and Gaurav Bahl$^\ast$\\
		\footnotesize{Department of Mechanical Science and Engineering} \\
		\footnotesize{University of Illinois at Urbana-Champaign, Urbana, Illinois 61801, USA}\\
		\footnotesize{$^\ast$ To whom correspondence should be addressed; bahl@illinois.edu} \\
	}
	
	\date{\today}
	
	\begin{abstract}
	 Opto-mechano-fluidic resonators (OMFRs) are a new platform for high-throughput sensing of the mechanical properties of freely flowing microparticles in arbitrary media. Experimental extraction of OMFR mode shapes, especially the acoustic pressure field within the fluidic core, is essential for  determining sensitivity and for extracting the particle parameters. Here we demonstrate a new imaging technique for simultaneously capturing the spatially distributed acoustic pressure fields of multiple vibrational modes in the OMFR system. The mechanism operates using microparticles as perturbative imaging probes, and potentially reveals the inverse path towards multimode inertial detection of the particles themselves.
	\end{abstract}
	\maketitle
	Optical and mechanical resonant modes are the basis for the design and implementation of nearly all sensor technologies.
	Micro-mechanical resonators are routinely employed for measuring forces \cite{Mamin2001,Gavartin2012} and inertial motion\cite{Bernstein1993,Krause2012}, and for quantifying the properties of fluids \cite{Kim2013,Gil-Santos2015} and particles \cite{Burg2007,Hanay2012}.
	The typical operating principle for such sensors is to measure the perturbation of resonant modes due to an analyte. For instance, the frequency shift of a mechanical resonator can be used to infer mass of a single microparticle bound to it\cite{Gil-Santos2010, Hanay2012}.
	The sensitivity of such devices depends on the location where the interaction between the analyte and the resonator takes place. For example, mass sensors are insensitive at displacement nodes and most sensitive at anti-nodes \cite{Liu2013,Olcum2015}.
	It is thus essential to map the resonant mode spatially, in addition to their spectral characteristics, in order to produce calibrated and well-optimized sensor devices.
	While several methods are available for imaging mode shapes of solid-state microdevices, including laser Doppler vibrometry \cite{Stanbridge1999,Lutzmann2016} and atomic-force microscopy \cite{Binnig1986,Meyer1988,Dareing2005}, there are relatively few techniques for mapping modes within fluids. Doppler imaging does permit visualization of acoustic pressure distributions in fluids \cite{Chan1987,Wiklund2007} but is limited to resolutions no better than 10's of \um \cite{Vakoc2009}. %
	Characterization of acoustic pressure distributions is thus necessary in resonant sensors in which fluids play a major role, especially for applications in biology and chemistry where fluid-based media are commonly encountered \cite{Burg2007,Kim2013,Bahl2013,Han2016}. 
	
	%
	%
	\begin{figure}[b]
		\centering
		\includegraphics[width=0.39\textwidth]{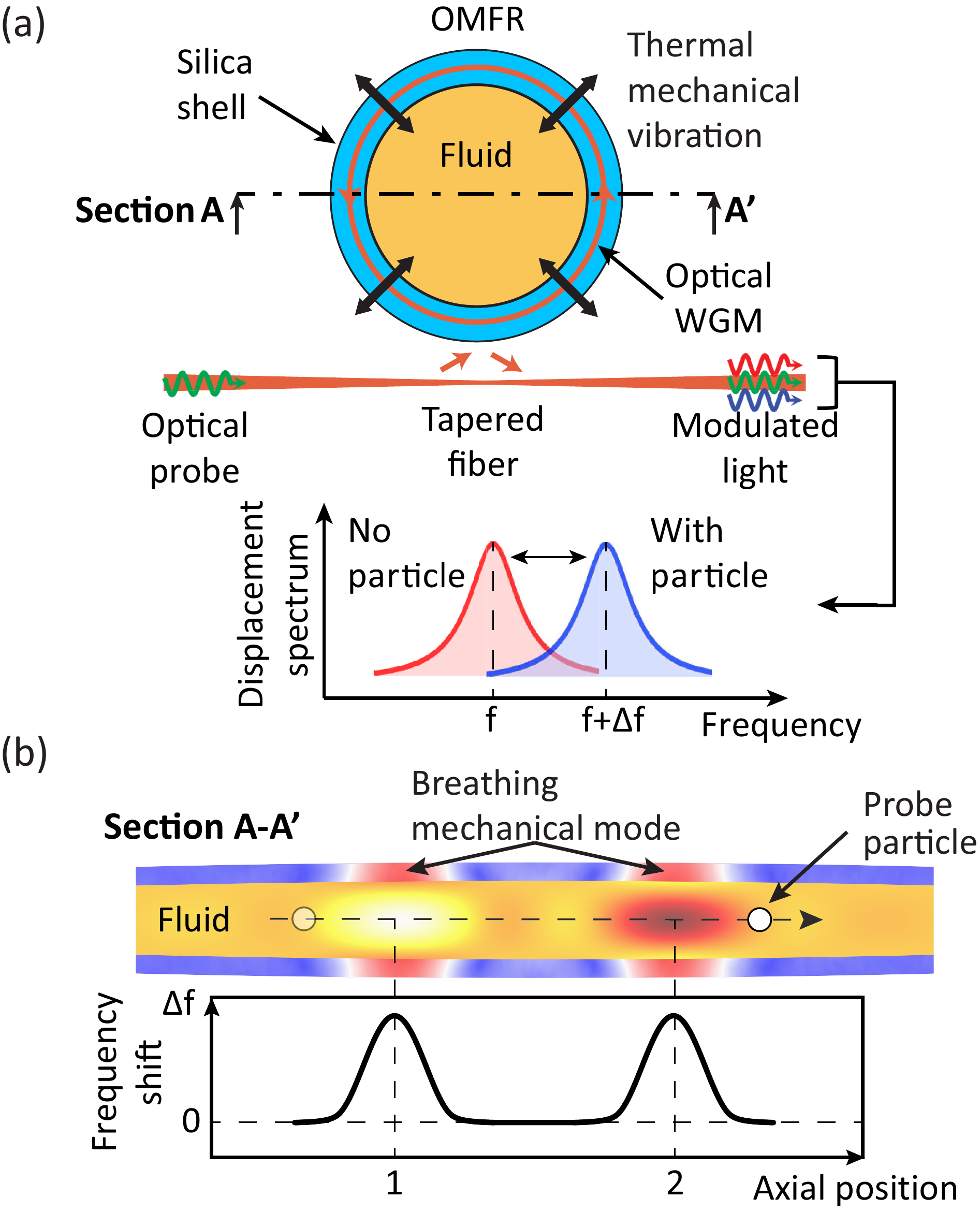}
		\caption{
			\textbf{Principle of particle detection and acoustic mode imaging in opto-mechano-fluidic resonators: }
			(a) A probe laser is coupled to optical whispering gallery modes (WGMs) through a tapered optical fiber. Thermal mechanical vibration of OMFR produces optical sidebands that are also coupled out through the waveguide. The spectrum of the mechanical motion is imprinted onto the modulated light.
			(b) Particles passing through the capillary interact with acoustic pressure distribution within fluid, causing mechanical frequency perturbation. This phenomenon allows us to image the acoustic pressure field within the fluid in the OMFR.
		}
		\label{fig:principle}
	\end{figure} 
	Previously, we have demonstrated opto-mechano-fluidic resonators (OMFRs) as a microfluidic sensor for measurements on bulk fluids\cite{Han2014_1, Kim2013, Bahl2013} and for determining the properties of individual microparticles\cite{Han2016} at extremely high speeds\cite{Suh2017}.	
	An example OMFR vibrational mode is illustrated in Fig.~\ref{fig:principle}(b), showing its hybrid nature in which both the mechanical strain of the shell and the acoustic pressure distribution within the internal fluid are coupled.	Specifically, the pressure field in the fluid forms a bridge between the optically measurable mechanical resonance on the shell and the properties of any analytes suspended in the fluid\cite{Han2016}. Thus, knowledge of the acoustic pressure field in the fluid can enable analysis of unknown particle properties such as compressibility, density, and volume. In this work, we introduce a technique for capturing the spatial shapes of the acoustic pressure modes of the constrained fluid in the OMFR system, using a single microparticle as a perturbative imaging probe.
	%
	%
	\begin{figure}[t]
		\centering
		\includegraphics[width=0.45\textwidth]{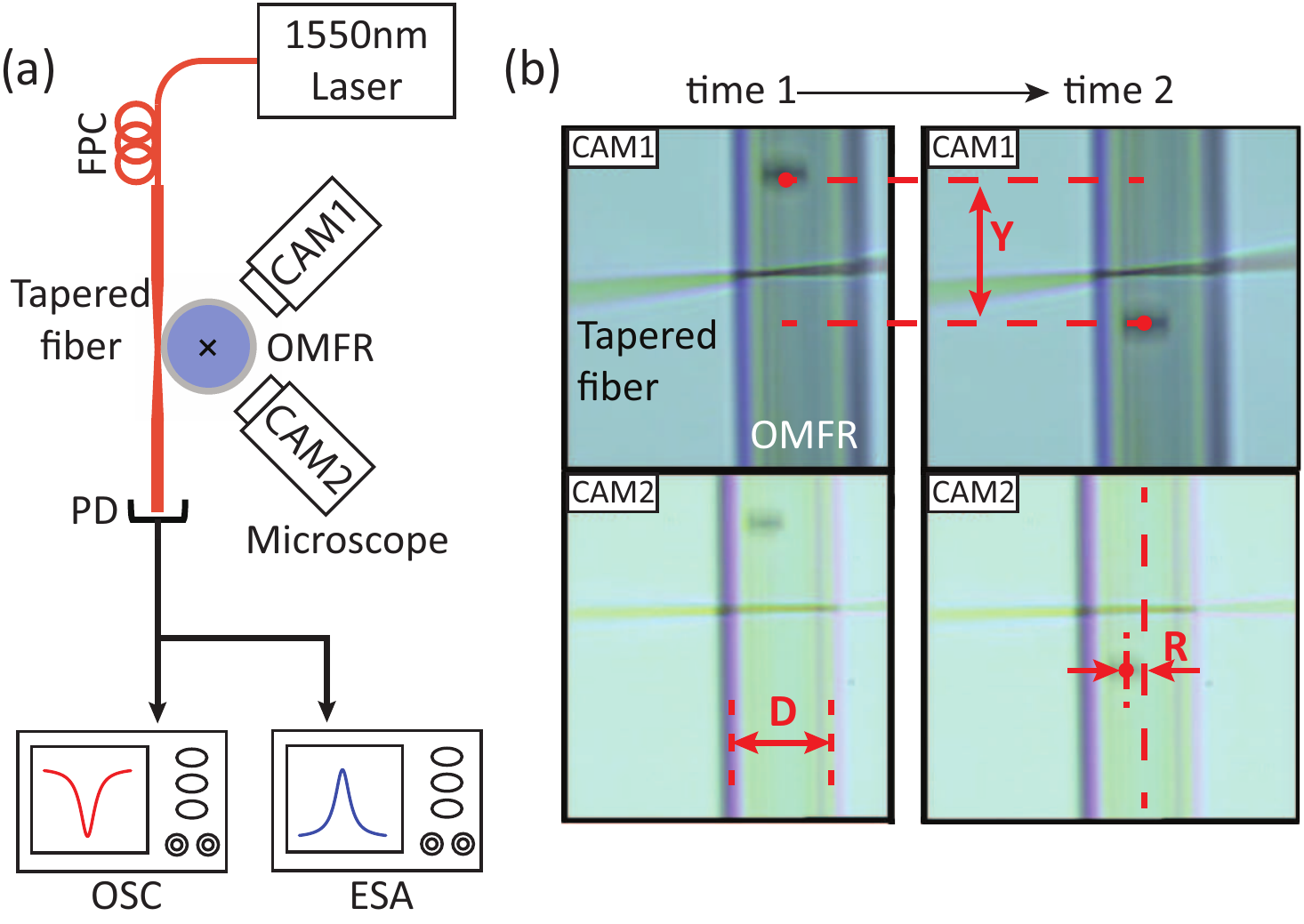}
		\caption{
			\textbf{Experimental setup:}  
			(a) Optical signals at the photodetector (PD) are converted to an electrical signal measured by an electronic spectrum analyzer (ESA) and an oscilloscope (OSC).
			Two cameras connected with microscopes are arranged perpendicularly for multiple viewing angles.
			(b) The axial movement (Y) and the radial position (R) of particles inside the OMFR and the diameter (D) of the OMFR can be measured using the cameras.
				}
		\label{fig:setup}
	\end{figure}

OMFRs are fabricated using a process previously reported in Ref.~\cite{Han2014_J}. In this work we produce OMFRs using 850 \um outer diameter fused-silica capillaries (Molex TSP-700850), which are pulled using linear actuators under CO\textsubscript{2} laser heating. During this process, periodic modulation of the laser power is used to pattern fixed diameter variations along the length of the microcapillary in the range of 40-60 \uum. High Q-factor optical whispering-gallery modes (WGMs) reside on the silica shell of the OMFR with primary confinement in the large diameter regions. A variety of mechanical modes (e.g. breathing modes and wineglass modes) are simultaneously confined at the same locations\cite{Bahl2012}.

Probe light is provided by a cw 1550 nm laser into an optical fiber and is coupled to the optical WGMs of the OMFR through a tapered region as shown in Fig.~\ref{fig:principle}(a).
Thermal-mechanical motion of the device at its mechanical resonance frequencies causes modulation of the optical resonance, thereby generating optical sidebands for the probe signal that carry the spectral information on the mechanical displacement\cite{Kippenberg2007,Aspelmeyer2014,Han2016}. The sensitivity to motion of this measurement technique is a function of the laser detuning from the optical resonance\cite{Aspelmeyer2014,Kippenberg2007}. Adjusting the detuning to an inflection point of the optical resonance thus produces the highest sensitivity to this mechanical motion. The mechanical vibration spectrum can thus be measured by beating the optical probe with these sidebands on a sensitive photodetector (PD), with the assistance of an electronic spectrum analyzer (ESA) as shown in Fig.~\ref{fig:setup}(a). Two microscopes image the OMFR at different angles and permit triangulation of the spatial position of any transiting particles. Particle velocity and OMFR geometry also can be measured from recorded videos as shown Fig.~\ref{fig:setup}(b).
%
%
\begin{figure*}[t]
	\centering
	\includegraphics[width=0.635\textwidth]{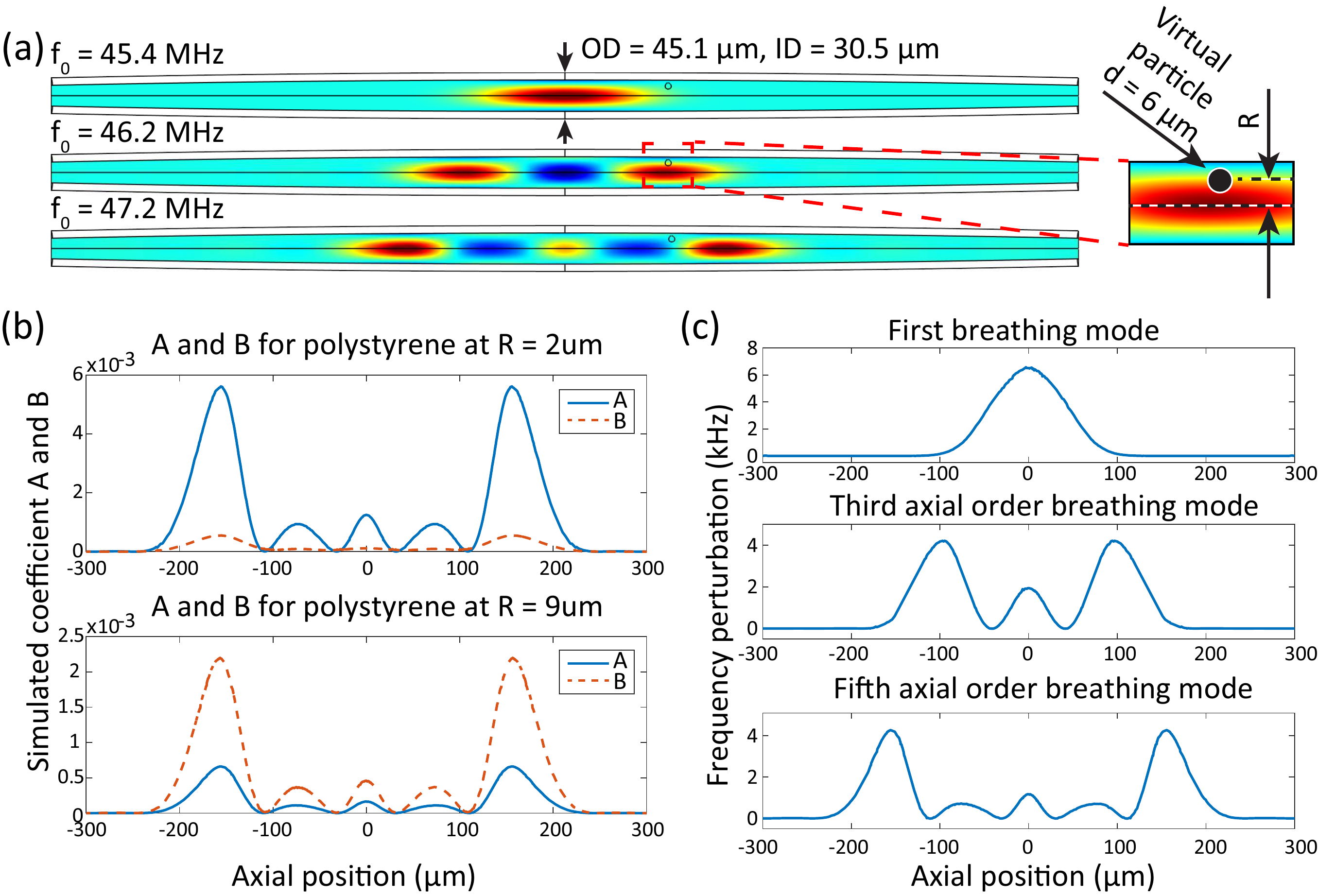}
	\caption{
		\textbf{Simulation of acoustic modes in fluid region and estimation of frequency perturbation:}
		(a) Multiple axial breathing modes are found in FEM simulation. The OMFR in simulations and experiments has 45.1 \um OD and 30.5 \um ID. A virtual particle is introduced to calculate the frequency shift using perturbation theory, and is placed at radial position (R) from the OMFR axis.
		(b) The coefficients of A and B in the perturbation theory are calculated for two different radial positions.
		(c) The predicted frequency perturbations reflect the acoustic pressure field in the fluid region, thereby enabling mapping.
	}
	\label{fig:simulation}
\end{figure*}
%
%
\begin{figure*}[]
	\centering
	\includegraphics[width=0.675\textwidth]{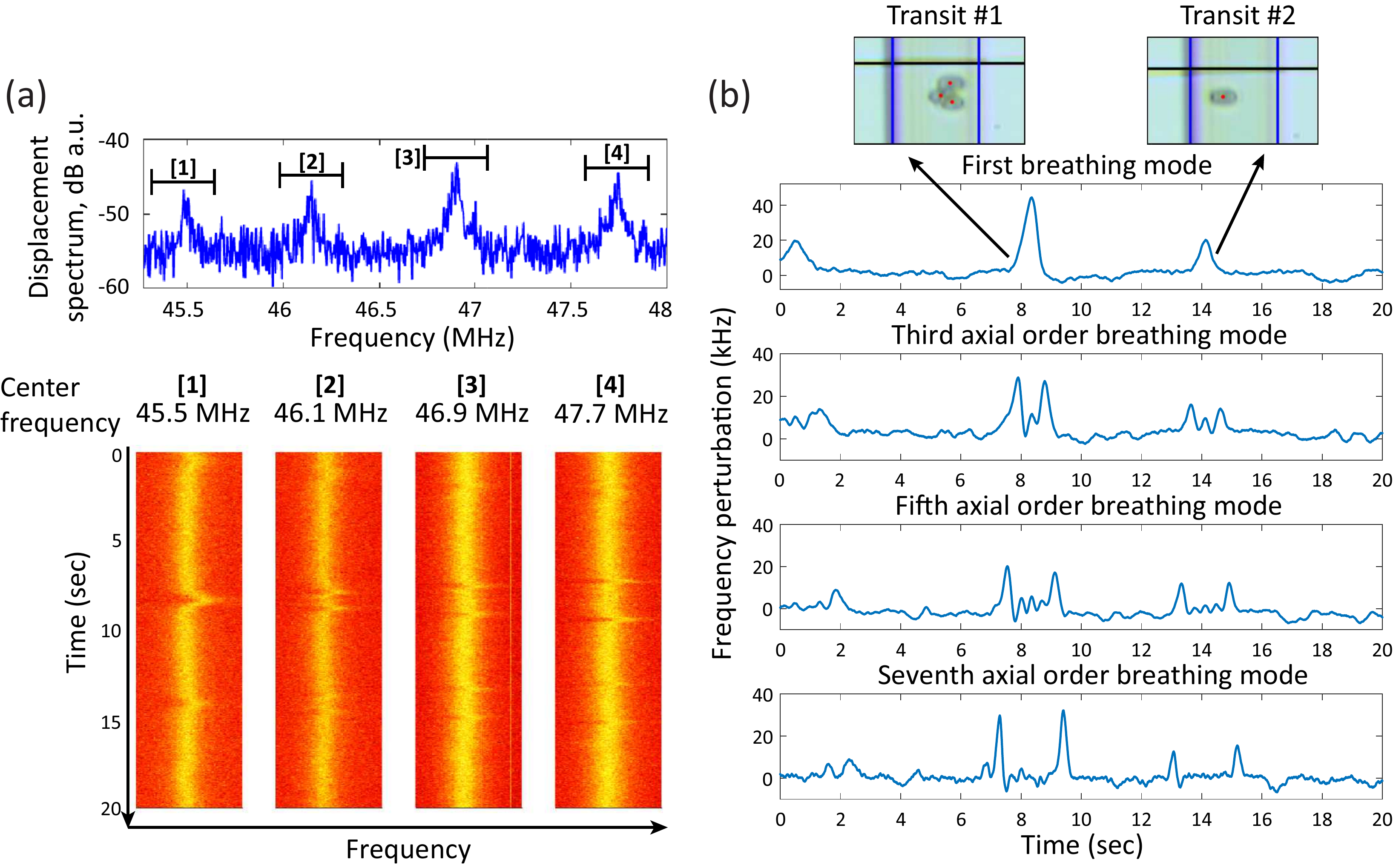}
	\caption{
		\textbf{Experimental observation of multiple mechanical modes and their particle sensitivities:}
		(a) (Top) Multiple breathing modes of the OMFR can be monitored using the ESA. (Bottom) The frequency fluctuation of each mode is tracked over time via spectrogram.
		(b) The center frequency tracks of each mode are reproduced in each plot. The frequency perturbations due to particle transits are visible at t = 8 s (transit 1) and 14 s (transit 2). Insets show camera images of the detected particles at the time when the frequency shifts occur. Blue and black lines in the insets indicate the edge of the capillary and the position of the tapered optical fiber respectively.
	}
	\label{fig:s_s}
\end{figure*}

	When a small particle transits through the resonator region of the OMFR, it increases both the effective stiffness and the effective mass of the hybrid fluid-shell mechanical modes. This occurs because solid particles typically have lower compressibility and higher density than that of water. The effect of such particles has been previously modeled via perturbation theory \cite{Han2016}. In this work we perform a linearization of the model from Ref.\cite{Han2016}, which is permissible in the limit of small particles. The following equations describe the frequency perturbation due to a particle of volume $V_p$, having density $\rho_p$, and compressibility $\kappa_p$.
	\begin{equation} \label{eq:1}
		\Delta f=\frac{1}{2}f_0 \left( \frac{\kappa_f-\kappa_p}{\kappa_f}A+\frac{\rho_f-\rho_p}{\rho_p}B \right)
	\end{equation}
	\begin{equation*} 
		\textstyle \text{where  } A= \frac{\int_{V_p} |p|^2dV}{\int_{V_f} |p|^2dV}\text{  and  } B= \frac{\int_{V_p} |\nabla p|^2dV}{k_l^2\int_{V_f} |p|^2dV}.
	\end{equation*}
	Here, $\Delta f$ and $f_0$ denote the frequency shift and original frequency respectively, and $p$ is the acoustic pressure distribution in the fluid fraction of the resonant container. $V_f$, $\rho_f,$ and $\kappa_f$ represent volume, density and compressibility of the carrier fluid and $k_l=2\pi \Omega_0/c_l$ is the unperturbed wavenumber associated with mechanical modes in the fluid fraction in which the speed of sound is $c_l$. The coefficients $A$ and $B$ correspond to the total acoustic potential energy and the acoustic kinetic energy over the particle volume respectively. 
	Since information on the acoustic pressure field in fluid region is embodied in $A$ and $B$, tracking frequency perturbation due to a single particle probe can be a way to map the acoustic modes of the constrained fluid in the OMFR system.

	In this experiment, we used a capillary OMFR having 45.1 \um outer and 30.5 \um inner diameter respectively, with water infused into the internal microchannel. Polystyrene microspheres of diameter d~=~6 $\pm$ 0.2 \um (Corpuscular C-PS-6.0) were used as imaging probes flowing through the device since they produce clear mechanical frequency perturbation during transits.
	Since each vibrational mode of the OMFR has a unique pressure distribution $p$, we must obtain $A$ and $B$ maps for each mode via Comsol Multiphysics simulation. To obtain these sensitivity parameters, a virtual particle is introduced in the simulation as shown in the inset of Fig.~\ref{fig:simulation}(a). The geometry of the simulation structure is confirmed by matching with four mechanical modes experimentally detected at 45.5, 46.1, 46.9, and 47.7 MHz, as presented in Fig.~\ref{fig:s_s}(a).
	Based on experimental image analysis in Fig.~\ref{fig:setup}(b), the single transiting particle is placed at a radial location of 9 \um from the center of the OMFR axis and a trajectory that is nearly parallel to the axis. Thus, we set the virtual particle at the same 9 \um offset, and having a trajectory parallel to the OMFR axis. Using this information, we are able to estimate both $A$ and $B$ for each of the simulated modes (example in Fig.\ref{fig:simulation}(b)) and $\Delta f$ as a function of particle axial position as shown in Fig.~\ref{fig:simulation}(c).
	In Fig.~\ref{fig:simulation}(b), we present two cases for the $A$ and $B$ calculation in which the virtual particle is placed at a radial location (R) of 2 \um and 9 \um from the center of the OMFR axis. The 2 \um case shows a trajectory that passes through the anti-node point in the pressure field such that the sensitivity to compressibility contrast (coefficient A) is higher than that of density contrast. On the other hand, the 9 \um case shows a higher sensitivity to density contrast (coefficient B). The calculated $A$ and $B$ in the simulation also clearly show the spatial positions of the anti-node and node points of the acoustic pressure field in the fluid corresponding to each mode.
	This dependence, as expected from the theory, indicates that acoustic pressure field spatially distributed in fluid region can be mapped by measuring frequency perturbation due to a probe particle.

	We can now experimentally confirm the acoustic pressure mapping based on the sequence of nodes and anti-nodes identified through the experimental frequency measurements using a single particle probe.
	Fig.~\ref{fig:s_s}(a) shows the measured spectrum and spectrogram in which four mechanical modes are simultaneously tracked over time during particle transits. Fig.~\ref{fig:s_s}(b) presents the fitted center frequencies of the four mechanical modes, extracted from this data, in which the particle transits can be clearly seen. Microscope images confirm the transits, revealing that the first signature in all data sets is produced by a cluster of particles (and is larger), while the second signature is produced by a single particle transit (and is smaller). Clearly, in order to obtain the best spatial resolution with a well defined probe, only single particle results should be collected.
	Using Fig.~\ref{fig:s_s}(b), we can also visualize the relative axial extent of these four acoustic pressure modes within the OMFR since all four signatures are produced simultaneously by the same probe particles. The lowest order mode exhibits a single anti-node as expected in Fig.~\ref{fig:simulation}(a). As the mode orders increase, the axial profile of each subsequent mode follows Fig.~\ref{fig:simulation}(a).
	%
	%
	\begin{figure}[t]
		\centering
		\includegraphics[width=0.4\textwidth]{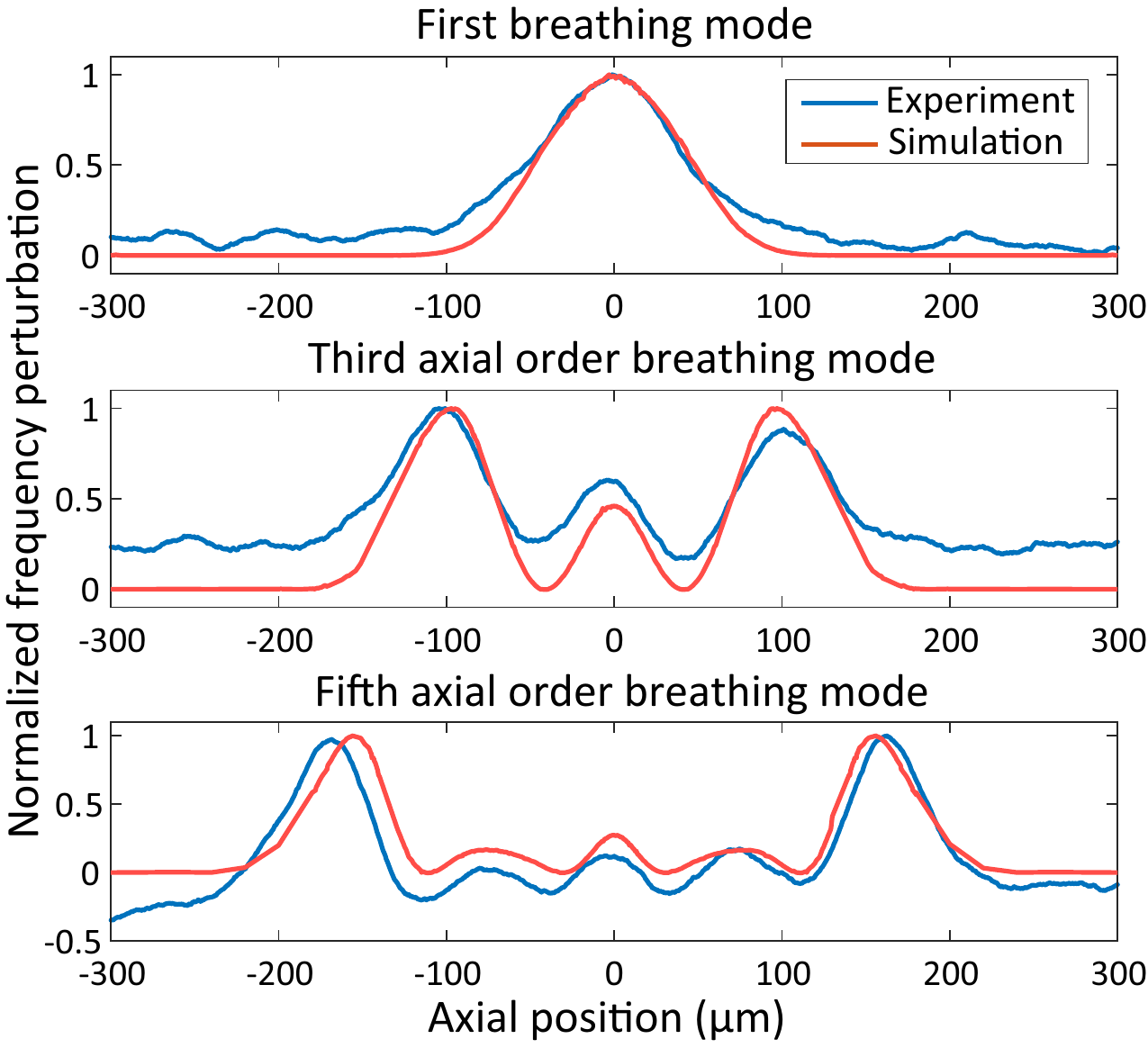}
		\caption{
			\textbf{Imaged spatial distribution of acoustic pressure field in fluid region: }
			Comparison between experimental result (Transit \#2 of Fig.~\ref{fig:s_s}) and simulation is shown. The x-axis of the experimental data is converted from time to length by using the velocity measured from the image analysis shown in Fig.~\ref{fig:setup}(b).
		}
		\label{fig:comparison}
	\end{figure}

	Upon comparing the predicted frequency shift with the experimental data, we find an underestimation of the calculated frequency perturbation. This discrepancy may arise from the following effects: first, the simulation employs a virtual particle and is reliant on the assumption that the particle does not affect the acoustic pressure distribution, which is not necessarily true in the experiment.
	Second, the analytical model is based on energy conservation\cite{Leung1982} but does not take account of loss due to damping, which could induce substantial difference for fluid-based resonators.
	Finally, the simulation estimations are based on correct triangulation of the radial position of the particles. However the estimation neglects the refractive index of the shell of the device, which is different from the fluid(i.e water). Because of this imaging distortion, the radial position is not perfectly measured, giving us a typical error bar in the triangulation of about $\pm$ 0.5~\uum.
	Since these factors primarily affect the predicted amplitude response, the mode shape can be clearly seen in a normalized comparison (Fig.~\ref{fig:comparison}). The simulated and experimental frequency-shift signatures are found to be in strong agreement.

	In the future, an improved analytical model supported by improved optical triangulation of particle position could improve the agreement between the experimental  and the simulated data. Additionally, hydrodynamic focusing that is employed routinely in other microfluidic systems\cite{Mao2012,Paie2014} can help eliminate the optical triangulation requirement and also permit repetition of the experiment for different probe particle trajectories, allowing reconstruction of the 3-dimensional mode shape.
	In the long-term, we envision that such multimode measurements\cite{Olcum2015,Hanay2015} could permit simultaneous extraction of multiple particle properties, such as density, compressibility, size, and position by using a well calibrated OMFR sensor.
\\
\par
	Funding for this research was provided through the National Science Foundation (NSF) Grant Nos. ECCS-1408539 and ECCS-1509391.

	\bibliography{mode_imaging_bib}

\end{document}